\documentclass[prl,aps,twocolumn,groupedaddress,floats,showpacs]{revtex4}
\usepackage{graphicx}
\usepackage{dcolumn}
\usepackage{bm}
\usepackage{color}
\def\he4{$^4$He}
\def\hel3{$^3$He}
\def\Am3{\AA$^{-3}$}
\def\beq{\begin{equation}}
\def\eeq{\end{equation}}
\begin{document}
\author{A. B. Kuklov}
\affiliation{Department of Engineering Science and Physics,
CUNY, Staten Island, NY 10314}

\author{L. Pollet}
\affiliation{Department of Physics, Arnold Sommerfeld Center for Theoretical Physics and Center for NanoScience, University of Munich, Theresienstrasse 37, 80333 Munich, Germany}

\author{N. V. Prokof'ev}
\affiliation{Department of Physics, University of Massachusetts,
Amherst, MA 01003, USA} \affiliation{Russian Research Center
``Kurchatov Institute,'' 123182 Moscow, Russia}

\author{B. V. Svistunov}
\affiliation{Department of Physics, University of Massachusetts,
Amherst, MA 01003, USA}
\affiliation{Russian Research Center ``Kurchatov Institute,''
123182 Moscow, Russia}


\title{Quantum Plasticity and Supersolid Response in Helium-4}

\begin{abstract}
We argue that the three key phenomena recently observed in solid \he4---mass supertransport, anomalous isochoric compressibility (syringe effect), and giant plasticity---are closely linked to each other through  the physics of an  interconnected network of tilted quantum-rough dislocations. As immediate implications of this connection several predictions follow: In the absence of \hel3 impurities, the syringe effect and giant plasticity persist down to $T=0$; the dynamical low-frequency syringe and giant-plasticity responses are dispersionless;  and similarly to giant plasticity but without direct relationship to the supertransport along the dislocation cores, \hel3 impurities should suppress the syringe effect partially or completely at appropriately low temperatures.
\end{abstract}



\maketitle

While the field of supersolidity in \he4 (for recent reviews, see, e.g., \cite{trends,Boninsegni_Prokofiev_review,Chan_Hallock_Reato}) is recovering from the 
``torsional-oscillator crisis,"~\cite{Beamish2012,Moses2014} it is important to realize that phenomena of 
disorder-induced supertransport \cite{shevchenko,PS,Hallock2009}, anomalous 
isochoric compressibility (syringe effect) \cite{Hallock2009}, and giant plasticity  \cite{Balibar},
discovered in the wake of the torsional-oscillator work, are interconnected and
are as interesting and non-trivial as supersolidity of perfect solids structures~\cite{Gross}. 
Theoretically, it is well established (using path-integral Monte Carlo simulations)
that ideal \he4 crystals are insulating~\cite{Ceperley,BPS,Clark,fate} and thus only long-lived
defects in the crystalline structure may be responsible for the  effects observed in Ref.~\cite{Hallock2009}.
The \he4 crystal thus emerges as a unique system demonstrating quantum behavior associated 
with the topological lattice defects. Here we observe that the experimental data strongly suggest that the quantum physics of the dislocation network in \he4 is not limited to superfluidity of dislocation cores~\cite{screw,stress,sclimb}.
The quantum-liquid state of dislocation kinks and jogs is likely to play a crucial part in the syringe  
and the giant plasticity effects at arbitrarily low temperature in the absence of \hel3 impurities. 
A low concentration of  \hel3 impurities brings to the forefront the fundamental but unsolved problem 
of the interaction between quantum-rough dislocations and impurities, which is
central to our understanding of the low-temperature limit.

The superfluid network idea proposed by Shevchenko~\cite{shevchenko} is strongly supported by 
numerical evidence that screw and edge dislocations with the Burgers vectors oriented along the 
$c$-axis~\cite{screw,sclimb} possess superfluid cores.
On the experimental side, a dc supertransport in solid \he4 has indeed been observed in a series of 
experiments with the ``UMass sandwich" setup \cite{Hallock2009}. In addition, an intriguing phenomenon 
of anomalous isochoric compressibility (syringe effect) has been discovered and attributed to the climb 
of edge dislocations assisted by supertransport along their cores (the so-called superclimb)~\cite{sclimb}. 
Very recently,  the same phenomena as in Ref.~\cite{Hallock2009}  have also been observed in the ``Alberta-ENS sandwich" setup \cite{Beamish_LT}, which is based on an inverted syringe effect, that is, inducing a superflow into or out of a solid by changing the external pressure on the latter.

The superfluid fraction estimates based on the observed dc mass flux (of the order of a few grams per year!) \cite{Hallock2009,Hallock2012} put the superfluid fraction at the level of  $10^{-10} -10^{-11}$ under
the assumption that the critical velocity is a few hundred m/s (as suggested by numerical simulations of the screw dislocation \cite{screw}). Such a fraction is orders of magnitude too small to be visible in torsional oscillator (TO) experiments.
The flow rate  as a function of chemical potential difference is reminiscent of the conductance anomaly
observed in 1D Luttinger-liquid systems \cite{Hallock2012,Hallock2014}. Overall, the dc flow experiments strongly support the picture of an interconnected network of dislocations with superfluid cores albeit with an unexplained temperature dependence of the flux. It is fair to say that these effects have no analogs in any other known material.

The giant plasticity effect has been observed in ultra pure and nearly perfect \he4 crystals~ Ref.~\cite{Balibar}.
The analysis of the solid \he4 shear response (see Ref.~\cite{Fefferman05}) is typically conducted within the classical vibrating string framework,  where  a dislocation is free to oscillate between  pinning  points created by the crossings with other dislocations as well as by  impurities.  This framework has been introduced originally in Ref.~\cite{Granato} to describe plasticity at relatively high temperatures  in commercial materials where dislocation motion is also strongly damped by bulk phonons and implies a
significant dissipation-induced dispersion (dependence on the frequency of the external drive).  Incidentally, the absence of this classical feature is often {\it interpreted} as key evidence for supersolidity in TO experiments \cite{Reppy2014}.

As will be discussed below, the classical dynamics of dislocations must completely freeze out at $T=0$ due to Peierls potential localizing kinks, so that, strictly speaking, the approach \cite{Granato} becomes inapplicable.  However, in the giant plasticity experiments the dislocations behave as if there were 
no Peierls barrier for the kink motion, raising fundamental questions about the role of quantum effects in the
dislocation dynamics.
We argue that solid \he4 is the only known substance where dislocations are exhibiting quantum behavior at low temperature. The quantum-liquid states of kinks and jogs lead to distinctively different effects: While the former is responsible for the giant plasticity in the absence of thermal activation and pinning centers, the latter 
is behind the syringe effect under the same conditions.

The quantum liquids of kinks and jogs share some 
similarities with each other and also with conventional 1D superfluids.
One common feature is the absence of dissipation, leading to an independence in the TO linear response on the frequency of the external force; {\it i.e.} the liquid groundstate of kinks \cite{Aleinikava2010} provides an alternative explanation to the claim of bulk supersolidity for the experimental results of Ref.~\cite{Reppy2014}. 

There is also a significant difference between both liquids: while the kink liquid does not need the core superfluidity for its existence, the quantum liquid of jogs can only exist in the presence of the core supertransport.

{\it Classical vs quantum kinks.}
In a finite-frequency setup, the criterion for quantum behavior involves a typical time required for a kink (jog) to tunnel through the periodic potential provided by the host lattice: this time has to be short compared to the experimental time scale. In conventional materials the Peierls barrier is too high for kinks and jogs to exhibit quantum motion. A classical kink on a gliding dislocation is usually viewed as the soliton solution of the sine-Gordon equation (see, e.g., Ref.~\cite{SG_dislocation}).
It is a particle with finite energy $E_k$ and extension $\lambda_k$ determined by the strength of Peierls potential, $u_p$, and the string tension energy, $\kappa$, both defined per unit length of the dislocation core. 
The kink mass $M_k$ is related to $E_k$ by the relativistic formula, where the role of speed of light is played by the speed of sound $c$. Exact relations between $E_k$, $\lambda_k$, $u_p$, $\kappa$ and the Burgers vector $b$
in the continuous limit $\lambda_k \gg b$ are
\beq
\lambda_k=\frac{b}{\pi} \sqrt{\frac{\kappa}{u_p}}\, , \qquad  E_k = \frac{4b}{\pi} \sqrt{u_p \kappa }\, , \qquad M_k=\frac{E_k}{c^2} \, ,
\label{size}
\eeq
In the continuous-medium approximation, the kink's motion along the dislocation is ballistic. In a real crystal, the discreteness of the host medium induces a periodic potential of some strength $u_{||} \propto u_p$ along the dislocation line with the period determined by the interatomic distance $a$ so that at zero temperature a classical  kink is trapped by the potential minima. It can only be moved by applying a finite external stress, which excludes the linear response regime. At large enough finite temperature, kinks move diffusively and their kinetics is controlled by the over-the-barrier activation time, which is exponentially long  by the factor $\sim \exp(u_{||}/T) \gg 1 $. Accordingly, the linear response of a dislocation on external stress can only be observed at very low frequencies.

The ratio $\lambda_k/a$  controls the ratio $u_{||}/u_p$. If $\lambda_k/a \gg 1$ (that is, $\kappa \gg u_p$), $u_{||}/u_p$ is exponentially suppressed by the factor $\sim \exp(-\lambda_k/a)$. In reality, however, both energy scales $\kappa$ and $u_p$ are of the same origin and, therefore, are of the same magnitude $\kappa \sim u_p$. Thus, $\lambda_k \sim a \sim b$ and $u_{||} \sim u_p$, which also implies $E_k \sim u_p$ and $M_k \sim m$, where $m$ is the atomic mass. We have ignored numerical coefficients of the order of unity in these relations.

Tunneling motion in solids depends on two energy scales: the interatomic interaction potential $u$ and the energy
of the atomic zero-point motion $E_0 \sim \sqrt{u \hbar^2/m a^2}$. In classical materials, $E_0 \ll u$. 
Conversely, in quantum crystals,  $ E_0 \sim u$. The semiclassical estimate for the tunneling rate between the potential minima, $\tau^{-1}=(E_0/h) {\rm e}^{-S}$ with   $S\approx \pi N$, where $N$ stands for the number of the energy levels under the tunneling barrier. This number scales as  $N \sim  u/E_0 \sim \sqrt{mu}\,a/\hbar $. As we discussed above, since kinks have sizes comparable to the interatomic 
distances, one can approximate
$E_0 \sim \hbar \omega_D$, where $\omega_D$ is the Debye frequency, and
$u \sim u_{||}a \sim u_pa$. If $S>50$  (in most materials $u_p$ is about a factor of $100$ larger than $E_0$) the tunneling time becomes infinite for all practical purposes.  This means that the kink's motion is arrested by the potential $u_{||}$ at low temperature $T\ll \omega_D$, unless there is an external bias driving the kink over the barrier.
In contrast, a kink characterized by
\beq
S \leq 30
\label{class}
\eeq
can tunnel through the barrier on the experimental time scale $t_{\rm exp}$ shorter than a second.
We shall call such a kink {\it quantum}.
In solid \he4, the two-body interaction at a typical inter-particle separation  $a \approx 3.5\,$\AA~  is 
$u \approx 10\,$K, smaller than the Debye frequency, leading to the tunneling action $ S$ close to unity.
This translates to tunneling times below 0.1ns, that is, seven orders of magnitude
shorter than typical frequencies employed in helium experiments.  

{\it Quantum glide.}
A dislocation totally confined to one Peierls valley has no kinks at $T=0$ and, therefore, cannot significantly contribute to plasticity. Long-range elastic forces between kinks through the bulk exclude the possibility of
having a quantum roughening transition \cite{norough}. At finite $T$, activated kink pairs lead to the 
dislocation contribution to plasticity $\sim \exp( - 2E_k/T)$.
Generically, however, the dislocation forest consists of tilted dislocations with multiple segments belonging to different Peierls valleys. Thus even at $T=0$ we have to deal with the finite concentration of kinks.
In classical materials this does not bring new physics as kinks remain immobile on experimental time scales.  
In quantum crystals, however, kinks can form a superfluid ground state  \cite{Aleinikava2010} as opposed to the 
insulating kink-crystal superlattice. This outcome is reminiscent of superfluid states obtained by doping Mott insulators with particles or holes. In other words, a tilted dislocation can still behave as a free string
and exhibit linear response upon applying external bias. We believe that it is this type of behavior that explains the giant plasticity observed at $T<0.3\,$K  in solid \he4  \cite{Balibar}.
 
Similarly, jogs on a tilted dislocation with superfluid core  may form a rough ground state ensuring that the syringe effect survives down to arbitrarily low temperature in impurity-free crystals. These arguments supersede the conclusion of Ref.~\cite{sclimb}, where it was assumed that each dislocation lies in a single Peierls valley, and thus remains smooth at $T=0$. It is important to note that, while the quantum liquid of kinks  is essentially a Luttinger liquid state (supporting sound propagation), the quantum liquid of jogs features infinite compressibility and thus a non-Luttinger-liquid, free-particle-like dispersion for collective excitations~\cite{sclimb}.

{\it Role of impurities}.
It is an experimental fact that both the giant plasticity \cite{Balibar} and the dc supertransport phenomena \cite{Hallock2014,Beamish_LT} are dramatically affected by \hel3 impurities  to the degree that both effects completely disappear at sufficiently high \hel3 concentration and low temperature.
A full theory of dislocation interaction with impurities, which includes quantum and thermal fluctuations of the dislocation, as well as the dynamics of impurities and the interaction between different dislocations forming the network, is yet to be developed. Meanwhile, it is important to note that taking into account the quantum nature 
of kinks while analyzing pinning and de-pinning processes, has no classical-physics substitute.  
The role of dislocation-shape fluctuations in pinning/de-pinning has been demonstrated in Ref.~\cite{Aleinikava2012}: At finite $T$ they induce a crossover from a purely elastic response at low $T$ to the giant plasticity regime at higher temperature. In contrast to the approach of Ref.~\cite{Fefferman05},
this crossover does not require dissipation (which is an additional feature on its own).

At the moment, the role of \hel3 in the syringe and superflow effects is not adequately understood. 
The most straightforward possibility for the superflow suppression observed at low temperatue \cite{Hallock2014}  
is blocking of the dislocation cores by the impurities (i.e., restricting the motion of \he4 atoms and jogs) ~\cite{Corboz08}.  
This is flow suppression at the {\it microlevel}. Dislocation pinning by impurities, works also 
at the {\it macrolevel}. Specifically, in the absence of impurities, a tilted superclimbing dislocation 
exhibits giant isochoric compressibility  $ \kappa_{\rm isoch} \propto L^2$, where $L$ is the length of the free dislocation segment. Pinning by \hel3 reduces $L$ down to a much smaller length scale $L_p$ determined by the density of the pinning centers along the core. This reduces compressibility and restores the Luttinger liquid nature of superfluidity in the core of the edge dislocation. Since the Luttinger parameter $K$ scales as 
$\propto \sqrt{\kappa_{\rm isoch}} \propto L_p$, it may fall below the critical value corresponding to
the Mott insulating state which results in the low-$T$ suppression of the superflow.
The true scenario can only be decided by further experiments or {\it ab initio} simulations.

{\it Summary and Outlook}.
All recent experiments~\cite{Hallock2009,Hallock2012,Hallock2014,Balibar,Beamish_LT} can be interpreted in 
terms of a network of quantum dislocations in \he4 that involves dislocations gliding and climbing 
in the basal planes.  The quantum behavior of dislocations is associated with two distinctively
different phenomena: the quantum-liquid state of kinks and superfluidity along the dislocation cores with the possibility of having the quantum-liquid state of jogs. The first one is responsible for the giant plasticity, 
the second one is behind the isochoric compressibility. In the absence of impurities, these effects most 
likely occur together at zero temperature, but the relationship between the two in superclimbing dislocations,
as well as the temperature dependence of the superflow flux observed in Ref.~\cite{Hallock2012}, remains unexplained. [The standard theory of quantum phase slips in 1D conductors predicts a power law dependence  $T^{-a}$ with $ a>0$ for the supercritical flow rate \cite{PhSl} at low temperature].
%

At low enough temperatures \hel3 impurities condense on dislocations and (i) pin them in space, and (ii) 
block supertransport along the cores. Theoretical and experimental studies of quantum dislocations, 
including a microscopic understanding of their interactions with impurities, is likely to bring new 
fascinating results. 

The exact nature of the weak dissipation at low temperature observed in Ref.~\cite{Balibar,Fefferman05}, which is ostensibly caused by the \hel3 impurities, is an open and important question. To answer it, one has to take into 
account the quantum nature of dislocations and consider \hel3 atoms as impuritons moving in a very narrow band~\cite{Kagan1974}. It should also be mentioned that, so far, there is no data on the dissipative 
effects in the syringe effect.

Compelling experimental confirmation of the proposed deep connection  between the three phenomena---giant plasticity, superclimb, and supertransport---requires further work. An important piece of
evidence might be the predicted persistence of the syringe effect down to arbitrarily low temperatures  \cite{remark} limited by the \hel3 concentration.

Comparing the $T$-dependencies of the shear modulus and the syringe response could be very informative.
Simultaneous ac measurements of the shear and pressure  responses may turn out to be crucial in revealing the relationship between the two. Such measurements can naturally be conducted in the setup of
Ref.~\cite{Beamish_LT}. Ac measurements may also help establish whether the supercritical dc flow can be transformed into the  ballistic ac superflow with the threshold on the applied pressure being proportional to the imposed frequency.

Although we have focused on quantum-rough dislocations, a quantum-rough grain boundary
is also a theoretical option (particularly, a grain boundary that can be viewed as an array of 
quantum-rough edge dislocations). As found in the simulations \cite{GB}, some grain boundaries in solid \he4 support 2D superfluidity. 
Thus, the mechanical response of such a granular medium can be viewed as a quantum analog of diffusive plasticity known as  Coble creep \cite{Coble}, where the high temperature diffusion along the grain boundaries is replaced by superfluid transport. Such a plastic behavior in response to superflow along grain boundaries, which we term {\it supershear}, should persist down to $T = 0$ and be dispersionless. A direct way to verify this effect is through the {\it inverse supershear} effect, that is, the superflow initiated by the shear deformation of a sample.

%
%
 
We are grateful to R. Hallock for his interest to this work and sharing unpublished data with us. This work was supported by the National Science Foundation under the grants PHY-1314469 and PHY-1314735, and by FP7/Marie-Curie Grant No. 321918 and FP7/ERC Starting Grant No. 306897.

\end{document}